\documentclass[twocolumn,showpacs,preprintnumbers,amsmath,amssymb]{article}

\usepackage{graphicx}
\usepackage{dcolumn}
\usepackage{bm}
\usepackage{amsmath}
\usepackage{color}
\usepackage{balance}
\usepackage{amssymb}
\usepackage{setspace}
\usepackage{url}
\usepackage{rsc}         
\begin{document}

   \renewcommand{\topfraction}{.9}
   \renewcommand{\bottomfraction}{.9}
   \renewcommand{\textfraction}{.1}
\title{Sensitive Chiral Analysis via Microwave Three-wave Mixing}

\author{David Patterson and John M. Doyle}

\date{\today}

\twocolumn[
  \begin{@twocolumnfalse}
    \maketitle
    \begin{abstract}
      We demonstrate chirality-induced three-wave mixing in the microwave regime, using rotational transitions in cold gas-phase samples of 1,2-propanediol and 1,3-butanediol.  We show that bulk three-wave mixing, which can only be realized in a chiral environment, provides a sensitive, species selective probe of enantiomeric excess and is applicable to a broad class of molecules. The doubly resonant condition provides simultaneous identification of species and of handedness, which should allow sensitive chiral analysis even within a complex mixture.

    \end{abstract}
  \end{@twocolumnfalse}
  ]

\maketitle
\vspace{0.1in}

Chirality plays a fundamental role in the activity of many biological molecules and in broad classes of chemical reactions. Despite this importance, to date no simple, sensitive, and species-specific determination of enantiomeric excess has been demonstrated. Spectroscopic methods for determining enantiomeric excess include optical circular birefringence (CB), circular dichroism (CD), vibrational circular dichroism (VCD), and Raman optical activity (ROA). All of these chiral analysis methods yield zero signal in the electric-dipole approximation\cite{nonlinliquids}. In contrast, the electric-dipole signal from sum-frequency generation can be non-zero in a bulk chiral environment\cite{threewavemixingliquid}. Sum-frequency generation (SFG) in the infrared and visible has been observed in samples of chiral molecules in solution. \cite{Shensumfrequency}. Doubly resonant SFG in both the infrared and microwave regime has been proposed but not observed\cite{shapirodemethyl,Hirotatripleres}. Our group recently demonstrated enantiomer-sensitive spectroscopy by combining a resonant microwave field with a strong adiabatically switched orthogonal non-resonant (DC) electric field\cite{meNature}.  In this Letter, we demonstrate true sum-frequency generation, a type of three-wave mixing, on a chiral sample in the microwave regime.
 We use two orthogonally polarized resonant applied electric fields to induce a third mutually orthogonal field at their sum frequency. The phase of this induced field changes sign with enantiomer, and its amplitude provides a sensitive, quantitative measure of enantiomeric excess.


The physical mechanism underlying our method arises completely from the Hamiltonian of an asymmetric top in an external electric field\cite{townes1975microwave}. A molecule's three rotational constants, $A$, $B$, and $C$, and the corresponding dipole moment component magnitudes, $|\mu_a|$,$|\mu_b|$, and $|\mu_c|$, determine the rotational energy levels of the molecule in an electric field.
  All allowed rotational transitions in such a molecule can be classified as purely \emph{a}-type,\emph{b}-type, or \emph{c}-type. The corresponding transition matrix elements $\langle \Psi_1|\hat{\mu}_x|\Psi_2\rangle$, $\langle \Psi_1|\hat{\mu}_y|\Psi_2\rangle$, and $\langle \Psi_1|\hat{\mu}_z|\Psi_2\rangle$ are proportional to $\mu_a$, $\mu_b$, or $\mu_c$, respectively\cite{gordyandcook}. The sign of any two of $\mu_a$, $\mu_b$, and $\mu_c$ is arbitrary and changes with the choice of axes; in contrast, the sign of the combined quantity $\mu_a\mu_b\mu_c$ is axis independent and changes sign with enantiomer.

\begin{figure}[h]
\includegraphics[width = 80mm]{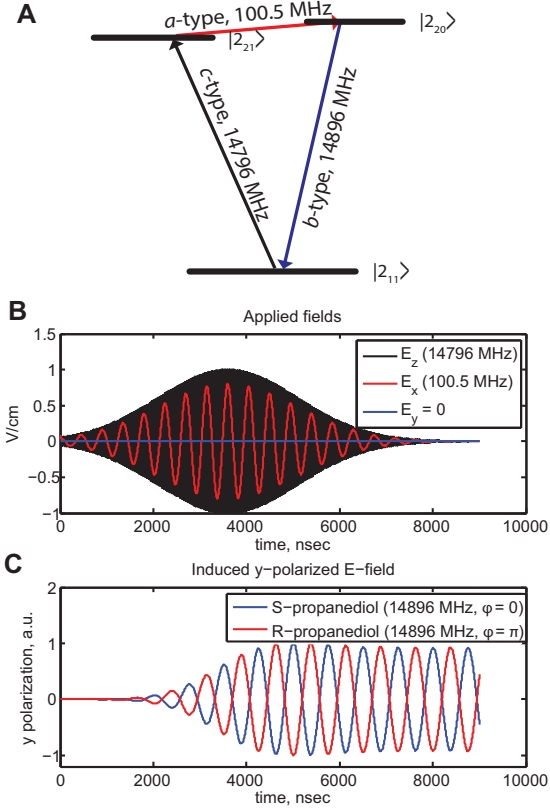} 

\caption{A: The relevant level structure for the demonstrated three-wave mixing in chiral 1,2-propanediol. B: Applied fields $E_z$ and $E_x$ used to induce enantiomer-dependent three-wave mixing. The fields shown are equivalent to the fields applied in the experiment. C: Predicted $\hat{y}$ polarized radiation. This prediction results from integration of the well-understood Hamiltonian of an asymmetric top in an electric field.
\label{theoryfig}}
\end{figure}

  The relevant rotational levels for 1,2-propanediol are shown in figure~\ref{theoryfig}A. We apply a $\hat{z}$ polarized electric field $E_c$ at frequency $\nu_c$ to drive a \emph{c}-type transition, and an $\hat{x}$ polarized electric field $E_a$ at frequency $\nu_a$ to drive an \emph{a}-type transition.  Applying pulses of these drive fields to the molecular sample induces $\hat{y}$ polarized radiation at frequency $\nu_b = \nu_a + \nu_c$; this field ($E_b$) emanates from a \emph{b}-type transition.  In the weak-pulse limit, $E_b$ is proportional to $\mu_a\mu_b\mu_c$, and for any strength of applied fields $E_b$ changes sign with enantiomer. Figure~\ref{theoryfig}C shows $E_b$ as predicted  from a simulation of the asymmetric top Hamiltonian of S- and R-1,2-propanediol. For an enantiopure sample the predicted amplitude of $|E_b|$ is comparable to the largest fields produced from strong rotational lines in ``traditional'' Fourier transform microwave spectroscopy (FTMW), while for a racemic sample $|E_b| = 0$.

  This enantiomer-dependent field $E_b$ is produced only by the fraction of the molecules initially occupying a single ro-vibrational state $|\Psi_1\rangle$ (see Figure~\ref{theoryfig}A). Molecules initially in the excited rotational states $|\Psi_2\rangle$ and $|\Psi_3\rangle$, connected to $|\Psi_1\rangle$ by pulses $E_a$ and $E_c$, produce a signal which interferes destructively with $E_b$.  At constant molecule density, the net signal $E_b \propto T^{-5/2}$, where $T$ is the rotational temperature of the sample. Cooling our molecular gas therefore greatly increases our signal. We use the buffer gas cooling method, developed previously and discussed elsewhere, to produce highly supersaturated gas-phase samples\cite{mefirstFTMW,meNature}.

\begin{figure*}[ht]
\includegraphics[width = 150mm]{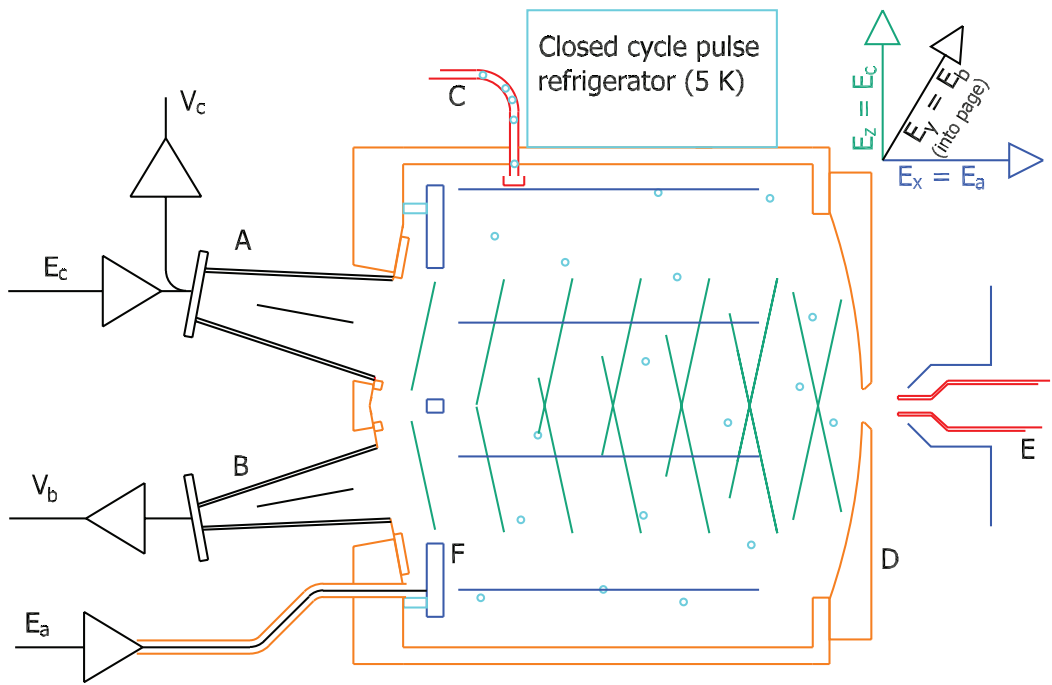} 
\caption{The experimental apparatus.  The cold molecular sample is injected into a cold (5.5 K) cell anchored to a closed cycle pulse tube refrigerator. Molecules are introduced to the cell through a warm (300K) injection tube E through a 1 cm aperture in the mirror D, and cold helium is introduced via a fill line C.
The resonant electric fields $E_c$ and $E_a$ are applied in two orthogonal directions ($\hat{z}$ and $\hat{x}$ in the naming convention of this paper), while the emitted radiation $E_b$ is polarized in the third ($\hat{y}$) direction.
$E_c$ is applied by broadcasting $\hat{z}$ polarized microwaves at frequency $\nu_c$ (green online) from horn A.  These microwaves are reflected from a curved mirror D and refocused onto a second microwave horn B that is oriented to collect $\hat{y}$ polarized microwaves. A fraction of the incident microwaves $E_c$ and the induced enantiomer-independent $\hat{z}$ polarized radiation are reflected back into horn A. A quasi-uniform electric field $E_a$ at frequency $\nu_a$ can be applied in the $\hat{x}$ direction  between the electrode F and the mirror D (blue online).   Horns A and B constrain $\nu_c$ and $\nu_b$ to be between 12 GHz and 18 GHz, and  $0 < \nu_a < 200$ MHz.
\label{apparatusfig}}
\end{figure*}

  Our apparatus is shown in Figure~\ref{apparatusfig}.  A pulse of microwaves is broadcast from a standard gain $\hat{z}$ polarized horn through the cold sample towards a spherical mirror.  The mirror reflects and refocusses the microwaves; 
  the combination of the incident and reflected beams forms $E_c$.  A second horn collects the induced $\hat{y}$-polarized microwave field $E_b$.
  Immediately in front of both horns is an electrode which can be independently charged. A voltage at frequency $\nu_a$ applied to this electrode produces the $\hat{x}$ polarized electric field $E_a$ (between this electrode and the spherical mirror).

A continuous stream of gas-phase 1,2-propanediol or 1,3-butanediol molecules (Sigma-Aldrich) enters the cold (5.5 K) cell from a warm (300 K) feed tube.  The molecules cool through collisions with the cold helium buffer gas.  The cold molecules remain in the gas phase for several milliseconds, until they diffuse to a cold cell wall, where they freeze. The gas phase molecules are subject to two simultaneous 3 $\mu$sec duration electric field pulses $E_c$ and $E_a$.   ${E_c}$  has strength $|E_c| \approx 1.6 $ V/cm, and has frequency $\nu_c = 14795.7$ MHz, tuned to the $|2_{11}\rangle \rightarrow |2_{21}\rangle$ transition in 1,2-propanediol\footnote{We use the $|J_{{k_{-1}}{k_1}}\rangle$ nomenclature of Townes and Schawlow\cite{townes1975microwave}}.
${E_a}$ has strength $|E_a| \approx 1.2$ V/cm, and has frequency $\nu_a$ = 100.5 MHz, tuned to the $|2_{21}\rangle \rightarrow |2_{20}\rangle$ transition.
$E_c$ and $E_a$  together resonantly drive molecules from the $|2_{11}\rangle$  state into a superposition $|\Psi\rangle = \alpha_1|2_{11}\rangle + \alpha_2|2_{21}\rangle + \alpha_3|2_{20}\rangle$, with enantiomer-dependent, complex coefficients $\alpha_i$.   When all external fields are turned off, the ensemble radiates with nonzero polarization $\vec{P}$ in all three polarizations.  $P_c$ ($\hat{z}$) oscillates at $\nu_{c}$, $P_a$ ($\hat{x}$) oscillates at $\nu_{a}$, and $P_b$ ($\hat{y}$) oscillates at $\nu_{b}$.  Molecules initially in the $|2_{21}\rangle$ and $|2_{20}\rangle$ states also radiate, with opposite phase, but a non-zero net field is produced due to the difference in thermal populations (about 15\% for our samples). $\vec{P}$ oscillates until the molecules rethermalize via collisions with background helium atoms, typically about $5$ $\mu$s. The emitted microwave fields $E_c$ and $E_b$, corresponding to $P_c$ and $P_b$, are collected by the microwave horns and amplified, producing voltages $V_c$ and $V_b$ respectively\footnote{The corresponding frequencies for 1,3-butanediol are as follows: $\nu_1 =$ 15052.2 MHz, $|3_{13}\rangle \rightarrow |3_{21}\rangle$; $\nu_2 =$ 168.0 MHz, $|3_{21}\rangle \rightarrow |3_{22}\rangle$; $\nu_3 = \nu_1 - \nu_2 = $ 14884.2 MHz, $|3_{22}\rangle \rightarrow |3_{13}\rangle$}.

Care must be taken to record $V_c$ and $V_b$ in a phase-repeatable way.  To this end, $E_c$ is generated by single-sideband-modulating a free-running carrier signal $S_c$ at frequency $\nu_{S_c} = $ 14760 MHz with a 35.7 MHz phase-repeatable waveform $F$, which is, in turn, generated by direct digital synthesis and is also used as a phase-stable trigger for our waveform recorder.  $E_a$ is generated by a gatable solid state amplifier driven by a free-running oscillator $S_a$ at frequency $\nu_a$.  $V_c$ is mixed with $S_c$ to produce a phase-repeatable, enantiomer-independent signal $V_i$ at $\nu_{c} - \nu_{S_a}$ = 35.8 MHz, while $V_b$ is mixed with $S_c$ and then with $S_a$ to produce a phase-repeatable, enantiomer-dependent signal $V_{\mathcal{E}}$ at $\nu_{c} - \nu_{S_1} - \nu_{S_2}$, also at 35.8 MHz.  A fast signal averager (Agilent U1084A) triggered on the first rising edge of the baseband waveform $F$ digitizes and averages the signal. The entire pulse sequence is repeated at 55 kHz, and many traces of $V_i$ and $V_{\mathcal{E}}$ are accumulated and averaged.  Similar signals $V_i$ and $V_{\mathcal{E}}$ were produced with $F$ replaced by a chirped waveform $F'$, resulting in a much broader polarization pulse $E_c$\cite{Patefirstbroadband}.
\balance

The sign of $V_{\mathcal{E}}$ is enantiomer dependent; in a racemic sample, induced radiation from S- and R- enantiomers is exactly opposite and the $|V_{\mathcal{E}}| = 0$. It is worth emphasizing that a racemic mixture will not radiate at the sum frequency $\nu_c + \nu_a$ despite essentially any geometric errors in the device, as three-wave mixing is strictly forbidden for a non-chiral bulk material under any conditions in the electric dipole approximation.  This definitively zero background for a racemic sample is a major advantage of this technique, making it particularly sensitive to slight enantiomeric excess.



\begin{figure}[h]
\includegraphics[width = 80mm]{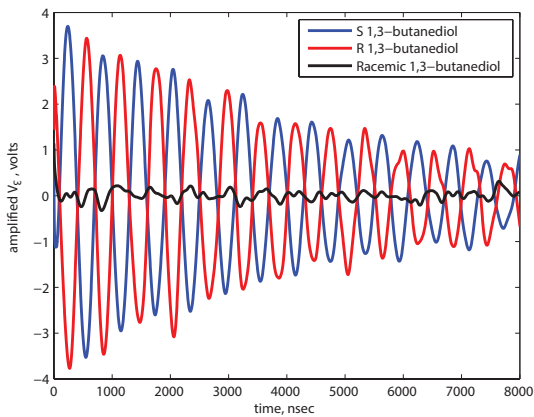} 
\caption{The digitized enantiomer-dependent signal $V_{\mathcal{E}}$ for an S-, R-, and racemic mixture of 1,3-butanediol.  The signal is emitted at 14884 GHz, but is digitized at an IF of 36.2 MHz and is shown here further mixed down to 4 MHz for clarity. As expected, the signal changes sign with enantiomer, while the racemic mixture shows no signal.
\label{glossydata}}
\end{figure}
\begin{figure}[h]
\includegraphics[width = 80mm]{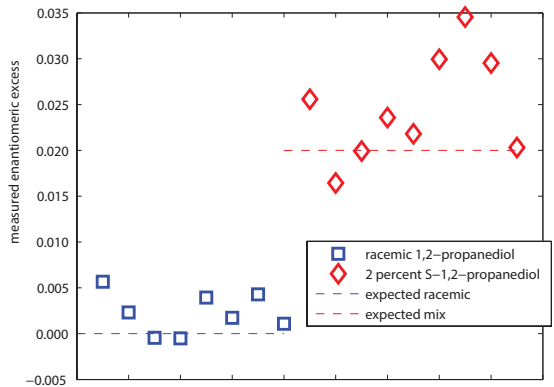} 
\caption{Repeated measurements of enantiomeric excess for a racemic mixture (blue squares) and a prepared 0.02 enantiomeric excess mixture (red diamonds). Each data point represents 5 million averages, consumes about 100 $\mu$g of sample, and uses 90 seconds of experimental time. The small but non-zero measured enantiomeric excess for the racemic 1,2-propanediol is believed to be an artifact resulting from trace amounts of S-1,2-propanediol adsorbed on the walls of the injection tube.
\label{highresfig}}
\end{figure}

Figure~\ref{glossydata} shows the radiated signal $V_{\mathcal{E}}$ for S-, R-, and racemic mixture of 1,3-butanediol.  As expected, the signal changes sign with enantiomer. Figure~\ref{highresfig} shows repeated measurements of a prepared 0.02 enantiomeric excess mixture and a racemic sample of 1,2-propanediol; the two samples are clearly resolved.  Our apparatus can resolve the difference between pure R- and S-1,2-propanediol in about 10 milliseconds, consuming about $10^{-7}$ grams of sample.


 An enantiomer-dependent signal $V_{\mathcal{E}}$ will only be produced efficiently if the appropriate phase-matching conditions are met such that the entire sample radiates constructively. This critical condition, which is often omitted
  in theoretical treatments, constrains the geometry of the apparatus and appropriate choice of transitions.  In our case, this condition is met by ensuring that $E_c$ and $E_b$ have a constant phase relationship over the entire sample, and the $\hat{x}$ polarized field $E_a$ is uniform over the entire sample.  This constrains the frequency of $E_a$ to $\nu_a \lesssim c/4L$, where $L$ is the characteristic length of the sample; a smaller apparatus could realize higher frequencies of $\nu_a$ at the expense of sensitivity\footnote{This limit is about 800 MHz for our $L$ = 15 cm cell, but parasitic capacitance between the electrode and the cell walls in fact limits us to $\nu_a \lesssim$ 200 MHz.}.
   In order to satisfy these conditions, a triad of electric-dipole connected states with two states nearly degenerate (in our case separated by less than 200 MHz) must be found.  The \emph{ideal} triad of states would have a very strong (observable) transition on one microwave transition and strong (drivable) transitions on the other two transitions.  This is the case with the triads used here ($|2_{11}\rangle$, $|2_{21}\rangle$, and $|2_{20}\rangle$ for 1,2-propanediol and $|3_{13}\rangle$, $|3_{21}\rangle$, and $|3_{22}\rangle$ for 1,3-butanediol).  It is not immediately evident that such states could be found for a generic chiral molecule.  We find, however, that many triads of states satisfying these conditions can be found for every molecule we checked (about 5), including carvone, limonene, and alaninol.  The dense microwave spectra of larger chiral molecules makes the existence of such triads essentially certain.  In addition, cold samples of these larger molecules typically exhibit several conformers with distinct constants $A$, $B$, $C$, $\mu_a$, $\mu_b$, and $\mu_c$, making the existence of an observable conformer with non-zero $\mu_a$, $\mu_b$, and $\mu_c$ extremely likely.


As a chemical analysis tool, the chirally sensitive microwave spectroscopy demonstrated here has several advantages over earlier linear and nonlinear  spectroscopy.  The vast majority of these spectroscopic chiral analysis techniques are performed in liquid samples with comparatively short coherence times. In contrast, the long coherence time of the gas-phase samples used here results in extremely narrow resonances, making the technique inherently mixture-compatible.  
In addition, the hardware required to produce, detect, and digitize the microwave fields used is significantly less complex than the corresponding ultrafast infrared laser systems used in infrared-visible SFG experiments.

The sum-frequency generation method used here is substantially more sensitive than our earlier enantiomer-specific microwave spectroscopy work. Furthermore, the apparatus used here contains no cavity and no moving parts, so the identity and enantiomeric excess of many individual compounds could be read out simultaneously.  This method has the significant advantage that a racemic mixture will not radiate at $\nu_b$ even in the case of unintentional crosstalk between the $\hat{z}$ and $\hat{y}$ polarized channels.  This is in contrast to our earlier work, where the enantiomer-dependent radiation was induced by an orthogonal DC field at the same frequency (but orthogonal polarization) to the enantiomer-independent radiation, analagous to $V_i$ here.

 Cooling of the molecules is essential to providing adequate signal. An alternative to the buffer-gas cooled source of cold molecules used here could be a pulsed supersonic jet\cite{balleflygare}.  Pulsed jets have recently been demonstrated on molecules as diverse as neurotransmitters, amino acids, and drugs\cite{Pate20121,Mata201291,nicotine}.  It would be straightforward to modify a pulsed-jet FTMW spectrometer to produce the electric field configuration used here, and our simulations suggest that R- and S-enantiopure samples of a broad class of molecules could be resolved in a single shot on such an instrument\cite{Patefirstbroadband}.

\noindent\textbf{Conclusion}
\balance
  A novel realization of three-wave mixing in cold gas-phase chiral samples has been demonstrated, providing a large-amplitude signal proportional to the sample's enantiomeric excess. This general technique promises to provide a sensitive, mixture-compatible chiral analysis technique.   Three-wave mixing was demonstrated here using molecular samples cooled via buffer gas cooling, and should also be easily detectable in cold gas-phase samples produced in pulsed supersonic beams.

  We thank Melanie Schnell and Dave DeMille for many helpful discussions; this work was funded by the United States Department of Energy.
\bibliographystyle{rsc}
\bibliography{chiralbib3}

\end{document}